\documentclass[prb,a4paper,twocolumn,floatfix]{revtex4-1}
\usepackage{color}
\usepackage[usenames,dvipsnames]{xcolor}
\usepackage{graphicx}
\usepackage{latexsym}
\usepackage[]{amsmath}
\usepackage{epstopdf}
\usepackage{amssymb}
\usepackage{array}

\usepackage{dcolumn}
\usepackage{bm}

\begin{document}
\title[AuBe - noncentrosymmetric superconductor]
{Fermi surface, possible unconventional fermions, and unusually robust resistive critical fields in the chiral-structured superconductor AuBe}

\author{Drew J. Rebar$^1$, Serena M. Birnbaum$^2$, John Singleton$^2$,
Mojammel Khan$^1$, J.C. Ball$^1$, P.W. Adams$^1$, 
Julia Y. Chan$^3$, D.P. Young$^1$, Dana A. Browne$^1$ and
J. F. DiTusa$^1$}

\address{$^1$Department of Physics and
Astronomy, Louisiana State University, Baton Rouge, LA 70803-4001,
USA\\ $^2$National~High Magnetic Field Laboratory, Los Alamos National
Laboratory, MS-E536, Los Alamos, NM 87545, USA\\ 
$^3$ Department of Chemistry, The University of Texas at
Dallas, Richardson, TX 75080}
\begin {abstract}
The noncentrosymmetric superconductor (NCS)
AuBe is investigated using a variety of thermodynamic and
resistive probes in magnetic fields of up to 65~T and
temperatures down to 0.3~K.
Despite the polycrystalline nature of the samples,
the observation of a complex series of
de Haas-van Alphen (dHvA) oscillations
has allowed the calculated bandstructure for AuBe
to be validated.
This permits a variety of BCS parameters describing
the superconductivity to be estimated, despite the complexity
of the measured Fermi surface.
In addition, AuBe displays a nonstandard field dependence of the
phase of dHvA oscillations associated with a band thought to 
host unconventional fermions in this chiral lattice. 
This result demonstrates the 
power of the dHvA effect to establish the properties of a single
band despite the presence of other electronic bands with a larger density 
of states, even in polycrystalline samples.
In common with several other NCSs,
we find that the resistive upper critical field
exceeds that measured by heat capacity and
magnetization by a considerable factor.
We suggest that our data exclude mechanisms for such an effect
associated with disorder, implying that 
topologically protected superconducting surface states
may be involved.
\end{abstract}
\maketitle

\section{Introduction}
Noncentrosymmetric superconductors (NCSs) have garnered much attention
over the past two decades; their lack of spatial inversion symmetry
breaks parity conservation via spin-orbit coupling, possibly resulting
in a mixed spin-singlet/spin-triplet superconducting pairing state
(see {\it e.g.}
References~[\onlinecite{Bauer:CePtSi,Yuan:LiPdB,Schnyd:topSC,Kaur:helicalSC}]
and references therein).  A spin-triplet state has been reported in
both strongly and weakly correlated materials such as CePt$_{3}$Si~[\onlinecite{Bauer:CePtSi}] and 
Li$_{2}$Pt$_{3}$B~[\onlinecite{Yuan:LiPdB}].  It is
possible that the superconducting phases of these materials are
topological~\cite{latenat}, supporting Majorana fermion surface
modes~\cite{Schnyd:topSC}.  In addition it has been predicted that
magnetic fields can induce a helical vortex phase in 
NCSs~\cite{Kaur:helicalSC}.  
Several NCSs
have crystal structures that also lack mirror symmetry, so that they
are better described as chiral-structured superconductors; these include
Li$_{2}$Pd$_{3}$B~[\onlinecite{latenat}], Li$_{2}$Pt$_{3}$B~[\onlinecite{Yuan:LiPdB}], 
BiPd~[\onlinecite{Sun:BiPd,khan19}], Mo$_{3}$Al$_{2}$C~[\onlinecite{Karki:MoAlC}], and
preliminarily RhGe~[\onlinecite{Tsvya:RhGe}].  In this context,
noncentrosymmetric AuBe is of great potential interest because of its
chiral crystal structure along with the presence of the heavy element,
Au.

AuBe forms in the $B20$ (or FeSi) crystal structure that has
attracted attention over the past decade because of the discovery that
magnetic materials that have this crystal structure, or that have the $P2_13$
space group, host skyrmion lattice states. Skyrmion lattices are
topologically stable field configurations with particle-like
properties~\cite{Muhl}.  Superconductivity in AuBe was originally
discovered in 1959 by Bernd Matthias~\cite{Matt:AuBe}, and
the material
has received more recent interest as a NCS~\cite{EteriPirate,Rebar}. In addition, materials having 
the $B20$ crystal structure have been predicted to host massless chiral
Fermions, motivating explorations of their electronic 
structures~\cite{bradlyn,tang,chang}. A recent report supports this 
identification in CoSi~[\onlinecite{takane}].
Thus, AuBe is an intriguing candidate 
material to search for unconventional superconductivity associated with 
its noncentrosymmetric crystal structure in combination with the possible existence of 
exotic quasiparticles. 

In this paper, we extend
the previous preliminary explorations to much higher magnetic fields $H$
and lower temperatures $T$.  Our AuBe samples
are exceptionally clean, so that, despite their polycrystalline nature, a plethora of de
Haas-van Alphen (dHvA) oscillations is observed at moderate-to-high
magnetic fields.  
These dHvA oscillations validate our electronic structure calculations,
allowing the Fermi
surface of AuBe to be deduced for the first time and
the density of states at the Fermi energy, vital for an understanding
of the superconductivity, to be derived. The
application of comprehensive magnetometry, resistivity and heat
capacity experiments at $^3$He temperatures has expanded the parameter
space of the superconducting phase diagram, permitting a Type I to
Type II crossover in the superconducting behaviour to be
observed. Furthermore, below the crossover, the $T\rightarrow 0$
resistive upper critical field is found to exceed that deduced from
magnetometry and heat capacity by a factor of around four, far beyond the 
expected critical field associated with a common superconducting 
surface state~\cite{StJames:surfSC}.
This large critical field, plus the observation of nonstandard dHvA oscillations,
may be associated with an electronic band in AuBe that is thought to host unconventional
fermions. 

This paper is organized as follows.  Section~\ref{Exp} covers the
 sample preparation, experimental techniques, and details of the
electronic structure calculations.  The normal-state properties, including the 
dHvA oscillations and their analysis, the calculated Fermi surface
 and the heat capacity are described in Section~\ref{Norm}, whilst
 Section~\ref{Sup} gives an account of the superconducting phase
 diagram.  A discussion of our findings and conclusions is given in
 Section~\ref{Disc}.

\section{Experimental and Computational Details}
\label{Exp}
Polycrystalline buttons of AuBe were synthesized by arc-melting
stoichiometric masses of high purity elemental Au (shot and wire) and
Be (chunks) in an Ar atmosphere.  In addition we found that small
single crystals $(0.2 \times 0.05 \times 0.05$~mm$^3$) formed in a
void of a large polycrystalline sample grown via modified Bridgman
growth technique employing a beryllium oxide crucible from United
Mineral \& Chemical Corporation.  The polycrystalline samples were
characterized by powder x-ray diffraction (XRD) on a Bruker D8 Advance
Powder Diffractometer equipped with a LYNXEYE detector.

The polycrystalline samples were cut via electric discharge machining
to an elongated bar shape and polished; they were then
characterized by heat capacity, dc magnetization, ac magnetic
susceptibility and resistivity measurements.  Heat capacity
measurements were performed in a Quantum Design (QD) PPMS system
equipped with a $^3$He insert.  Magnetization and ac magnetic
susceptibility measurements were carried out in a QD MPMS XL7.  The
identification of bulk superconductivity in AuBe, as well as the
values of the critical fields and temperatures, were verified by
magnetization experiments carried out on powdered arc-melted samples
and one tiny single crystal~\cite{tiny}.  Low temperature ac magnetic
susceptibility was performed within a Janis $^3$He insert at a
frequency of 19~Hz, employing a home-built susceptibility coil set
consisting of a primary drive coil and two series counter-wound
secondary pickup coils.  The real part of the ac susceptibility was
normalized at 1.8~K to the value reported by the MPMS at 1.8~K.
Resistance and magnetoresistance measurements were carried out on
rectangular shaped samples with electrical contacts formed via Epotek 
silver epoxy and thin platinum wires. These measurements
employed standard four-probe ac lockin techniques at 19~Hz, both in
the MPMS and in the Janis $^3$He insert. 

For the pulsed-field dHvA experiments, polycrystalline
needles were inserted into a 0.5~mm bore, 1.5~mm long compensated-coil
susceptometer, constructed from 50-gauge high-purity copper wire.  The
coil is wound with approximately 610 turns in one sense, followed by
around 390 in the opposite sense; final turns are added or subtracted
by hand on the bench-top to reduce the uncompensated area of the coil
to a fraction of a turn~\cite{HoNJP}. Fine-tuning of the compensation
is accomplished by electronically adding or subtracting a small part
of the voltage induced in a coaxial single-turn coil wound around the
susceptometer~\cite{HoNJP}.  Once this has been done, the signal from
the susceptometer is $V \propto ({\rm d} M/{\rm d} t) = ({\rm d}
M/{\rm d}H)({\rm d}H/{\rm d}t)$, where $M$ is the magnetization of a
sample placed within the bore of the coil and $H$ is the applied
magnetic field~\cite{HoNJP}.  Magnetic fields were provided by a
65~T, capacitor-bank-driven pulsed magnet at NHMFL Los Alamos with a
rise time to full field of about 10~ms and a downsweep time of about
80~ms (see Figure~2 of Ref.~\onlinecite{HystLoops}).  The susceptometer was
placed within a simple $^3$He cryostat providing temperatures down to
0.4 K. Magnetic fields were deduced by integrating the voltage
(proportional to ${\rm d} H/{\rm d}t$) induced in an eleven-turn coil
($\dot{B}$ coil), calibrated by observing the dHvA
oscillations of the belly orbits of the copper coils of the
susceptometer~\cite{HoNJP}.  A quantity proportional to the
differential susceptibility ${\rm d} M/{\rm d} H$
can be obtained by dividing the 
$({\rm d}M/{\rm d} t)$ signal by the $\dot{B}$-coil voltage.

\begin{figure}[htbp]
\centering
\includegraphics [width=0.98\columnwidth]{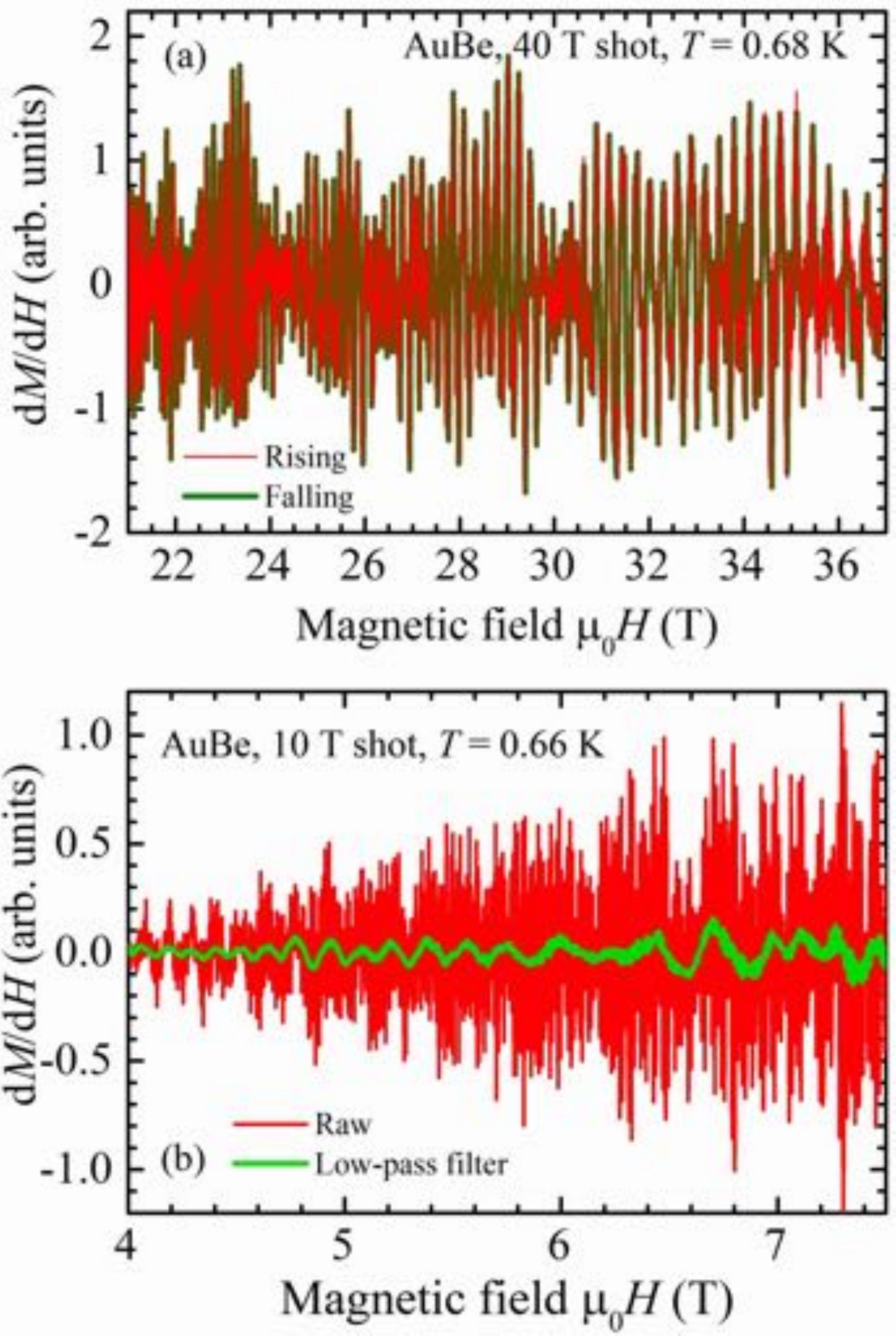}
\caption{(a) The differential susceptibility, ${\rm d}M/{\rm
d}H$, plotted versus magnetic field $\mu_0H$ for a polycrystalline AuBe rod at
$T=0.68$~K, recorded during a 40~T pulsed-magnet shot.  Several
different series of dHvA oscillations are visible. Data
for both rising and falling fields are shown.  The ``fur'' on the data
is not noise, but comprises the ``belly'' orbit dHvA
oscillations from the copper of the susceptometer coil.  (b)~The red
trace shows dHvA oscillations recorded using a 10~T
pulsed magnet shot ($T=0.66$~K) where the sample is the same as that used
in (a).  The green curve represents the application of a low-pass
filter to the red trace; this removes the higher-frequency de Haas-van
Alphen oscillations so that the lower-frequency series can be seen
more readily.  }
\label{dHvA}
\end{figure}

Electronic structure calculations were performed using the
WIEN2K~[\onlinecite{Blaha:Wien2k}] LAPW density functional software package,
using the Perdew-Burke-Ernzerhof~\cite{Perd:GenGrad} GGA functional.
The experimental lattice constant (see Section~\ref{struc}) was used.
The Au atom was placed at (\emph{u},\emph{u},\emph{u}) with
\emph{u}=0.844, and the Be was positioned at \emph{u}=0.154.  The
muffin tin radii used were 2.50 a.u. for Au and 1.90~a.u. for Be. The
plane wave cutoff in the code was varied from \emph{R*K}=7.0 to 8.5 to
ensure convergence~\cite{convnote}, and a 27$^{3}$ grid was used for Brillouin zone
integrations, which resulted in 654 points in the irreducible
zone. Calculations were performed both omitting and including the
spin-orbit interaction. This showed that the spin-orbit interaction 
makes little difference to the overall Fermi-surface
topology, but causes
an obvious, but small, splitting of the electronic bands except at high symmetry 
points of the Brillouin zone. 
The Fermi surfaces were rendered on a denser grid of 34$^{3}$ points.
(For clarity, the theoretical Fermi-surface sections shown in Figure~\ref{FS} below
are plotted for the case of zero spin-orbit interactions.)

\section{Normal-state properties}
\label{Norm}
\subsection{Structural details}
\label{struc}
The room temperature powder X-ray diffraction measurements confirmed
the $B20$ crystal structure (also known as the FeSi
structure)~\cite{FeSi}.  The underlying lattice is simple cubic with
four non-equivalent formula units per unit cell, and a lattice
constant (unit cell edge) of 0.4659~nm, in good agreement with earlier
work~\cite{EteriPirate,Cull:AuBe}.  The diffraction measurements also
identified small amounts of Au$_{2}$Be and BeO in our polycrystalline
samples, neither of which is a known
superconductor~\cite{handbook,Greenwood}.  A single crystal~\cite{tiny} was
characterized via X-ray diffraction which confirmed the $B20$
crystal structure but identified a high density of twin boundaries.

\subsection{deHaas-van Alphen frequencies}
\begin{figure}[htbp]
\centering
\includegraphics [width=0.96\columnwidth]{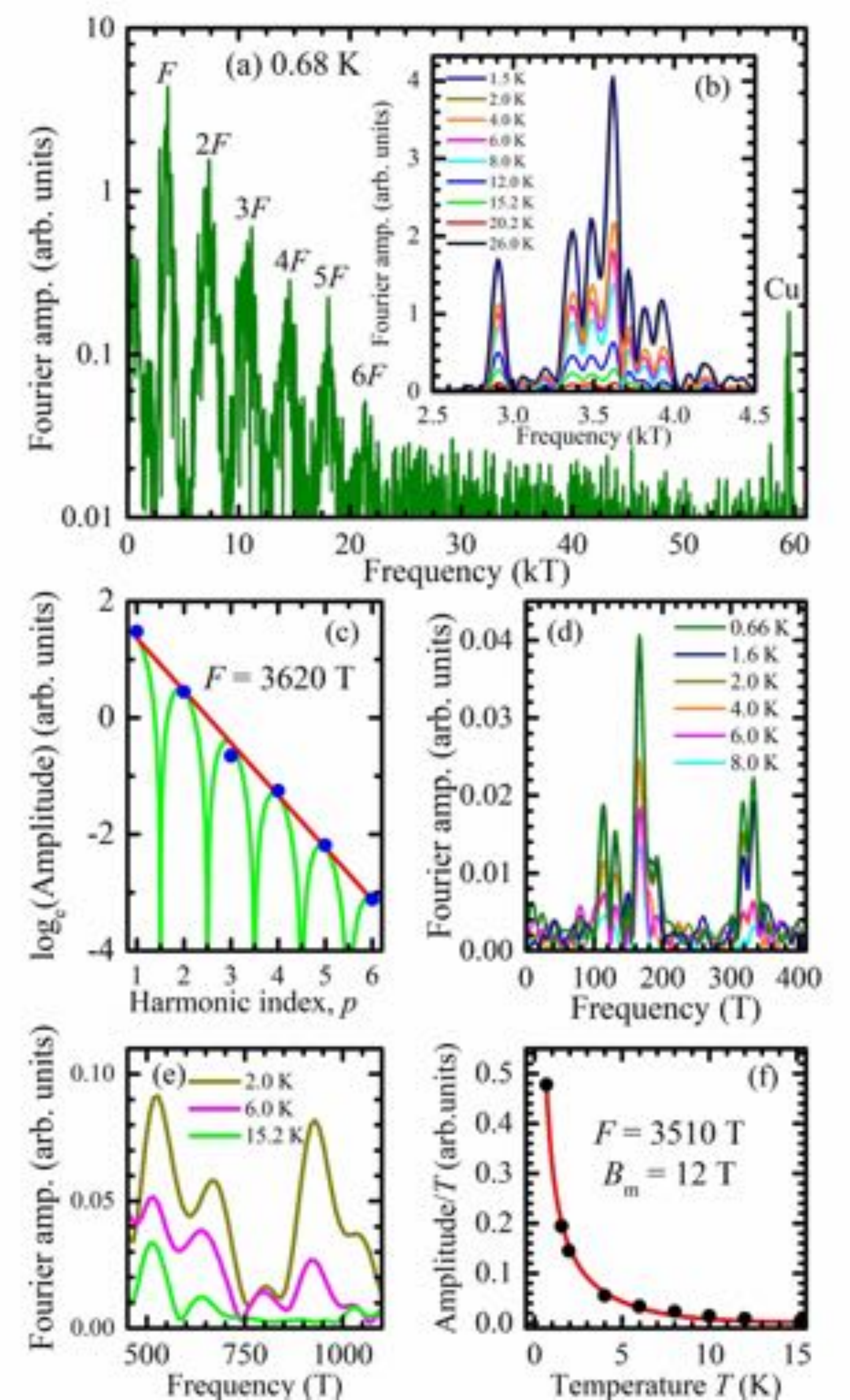}
\caption{(a)~Fourier spectrum ($20-38$~T window) of the data in
Figure~\ref{dHvA}(a) plotted with a logarithmic amplitude scale.  A
band of several series of dHvA oscillations (labelled
$F$) and their higher harmonics (labelled $2F.....6F$) are clearly
visible, along with a peak at 59.5~kT due to the belly orbits in the
Cu coil of the susceptometer.  (b)~[Inset]~Fourier transforms (linear
amplitude scale) of data from 40~T pulses recorded at a series of
higher temperatures $T = 1.5 - 26$~K. The same field window as in (a)
has been applied.  The frequency range has been chosen to show just
the fundamental frequencies in the band of oscillations labelled
$F$ in (a). (c)~Logarithmic amplitudes (points) of the
harmonics of the $F=3620$~T series of dHvA oscillations
versus harmonic index $p$;
the red line is a guide to the eye, showing an approximate linear
decrease with $p$. The green curve is
a fit of Equation~\ref{dongle} for $gm^*=2$. 
(d)~Fourier transforms of dHvA oscillations
from a series of 10~T pulsed-magnet shots at temperatures
$T=0.66-8.0$~K.  (e)~Field-axis expansion of Fourier transforms (linear
amplitude scale) of data from 40~T pulses for three example temperatures,
showing frequencies in the range $490-1200$~T. The same field window as in (a)
has been applied. (f)~(Fourier~amplitude)$/T$ versus temperature for the
$F=3510$~T series of dHvA oscillations. The data window
was centred on $B_{\rm m} = 12$~T.
The red line is a fit of Equation~\ref{LKM} to the
data, yielding $m^*=0.58\pm0.04m_{\rm e}$. }
\label{FFT}
\end{figure}

The differential susceptibility $({\rm d} M/{\rm d} H)$ measured for
both rising and falling magnetic fields using a pulsed-field shot with
a maximum field of 40~T is shown in Figure~\ref{dHvA}(a). The fact
that oscillations in both rising- and falling-field data overlay very
well shows that there is little or no inductive heating due to the
pulsed magnetic field~\cite{HeatingComment}.  Perhaps surprisingly,
given the polycrystalline nature of the samples, a plethora of de
Haas-van Alphen oscillations of several different frequencies $F$ is
observed.  The dominant oscillations at low temperature and high
fields have $F\approx 2900-4200$~T (see Figure~\ref{FFT}).
Figure~\ref{dHvA}(b) shows that these oscillations persist down to
fields of a few Tesla. In addition, the application of a low-pass
filter (green curve) reveals that low frequencies $F \approx 100-
200$~T are also present, in agreement with measurements of $M(H)$
performed at 1.8~K and fields of up to 7~T 
in a SQUID magnetometer~\cite{Rebar}.

The emergence of oscillations with frequencies $F\approx 4000$~T
at a field of about $4-5$~T [Figure~\ref{dHvA}(b)], gives an estimate of the lengthscale 
of the disorder encountered by the quasiparticles in the
polycrystalline AuBe samples.
The cyclotron radius, $l_{\rm c}$, is the characteristic size
of the orbitally quantized wave function and is given by
\begin{equation}
l_{\rm c} = \left(\frac{(2l_{\rm LL}+1)\hbar}{eB}\right)^{\frac{1}{2}},
\label{Gove}
\end{equation}
where $l_{\rm LL}=F/B$ is the Landau-level index and $B$ is the magnetic flux density~\cite{fieldnote,Shoenberg,AuZn}.
Inserting $B=5$~T and $F=4000$~T yields $l_{\rm c}\approx 0.46~\mu$m.
Optical microscopy of our polycrystalline AuBe samples indicates
grain sizes spanning the range $\sim 1-50~\mu$m.
Therefore, one possible explanation for the low-field onset of the dHvA oscillations in
Figure~\ref{dHvA}(b) is that when the magnetic field
exceeds $\approx 4-5$~T, the cyclotron radius becomes small enough
for the Landau wave functions to fit comfortably within even the smallest grains,
so that dHvA oscillations emerge and their amplitudes begin to
follow the Lifshitz-Kosevich formula~\cite{Shoenberg} with a constant scattering
rate determined by impurities within the grains~\cite{AuZn}.
As the field is lowered below $\approx 4-5$~T, the cyclotron radius grows
to the typical size of the grains and
increasing numbers of quasiparticles encounter the
boundaries, causing the scattering
rate to increase and the oscillations to vanish.

Figures~\ref{FFT}(a,b) show the Fourier spectra ($20-38$~T window) of
the data in Figure~\ref{dHvA}(a) and similar pulses recorded at higher
temperatures.  As mentioned above, the spectra are dominated by
several series of dHvA oscillations with frequencies
spanning the range $F\approx 2900 - 4200$~T [Figure~\ref{FFT}(b)],
plus their harmonics $2F, 3F\dots 6F$ [Figure~\ref{FFT}(a)].  The
presence of the higher harmonics is suggestive of exceptionally
sharply defined Landau levels, due to low quasiparticle scattering
rates~\cite{Shoenberg}.

\begin{figure*}[htbp]
\vspace{-25mm}
\centering
\includegraphics [width=0.75\textwidth, angle =90 ]{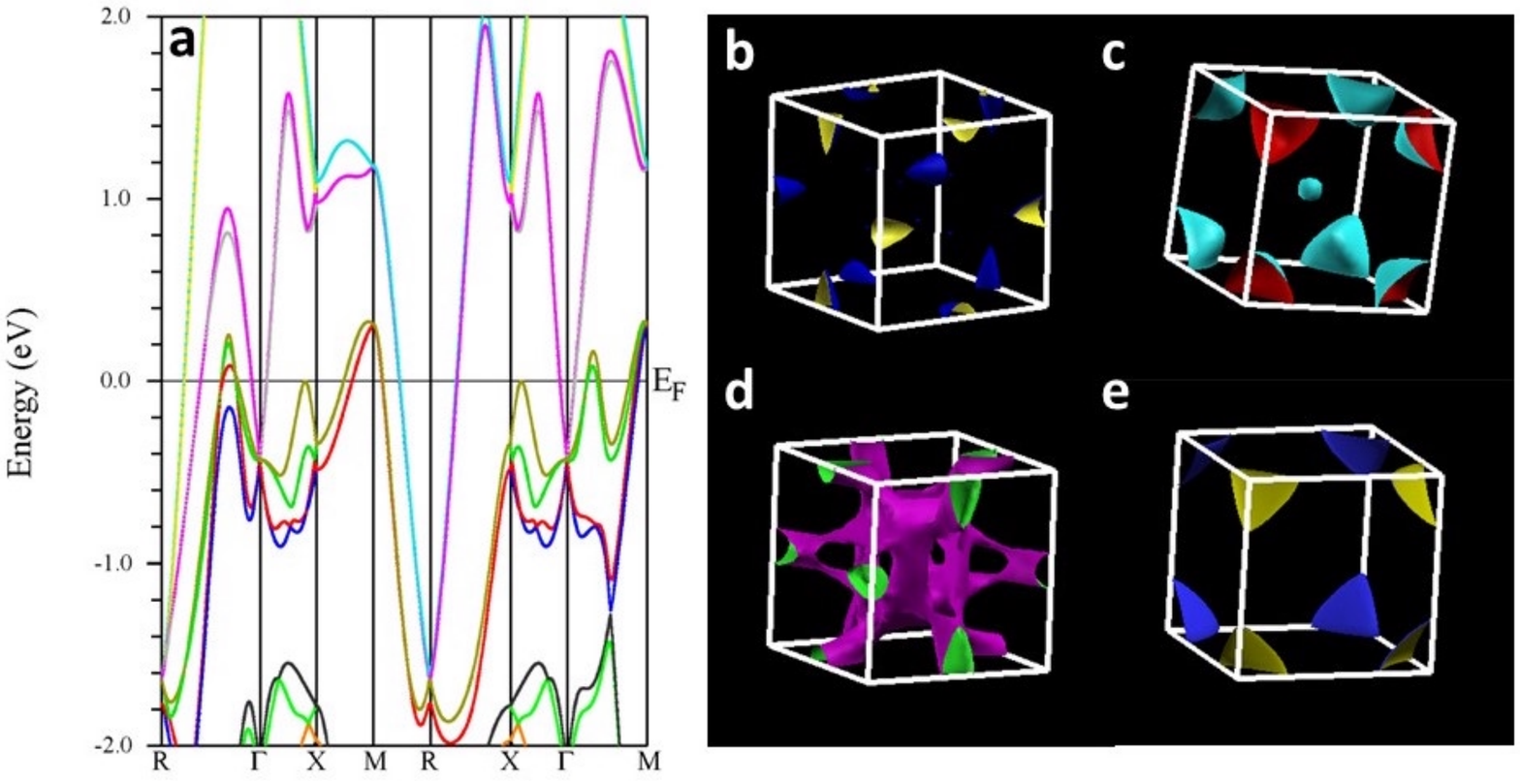}
\vspace{-30mm}
\caption{(a)~ Calculated electronic structure of AuBe including
spin-orbit coupling and (b-e)~the predicted Fermi-surface sections
shown within the simple-cubic Brillouin zone.
For clarity, the predicted Fermi pockets are shown here without the
slight doubling of surfaces due to band splitting 
caused by the spin-orbit coupling. 
(b)~Hole ellipsoids centred on the zone-edge
M-points; (c)~An electron approximate superellipsoid centred on the zone-corner
R-point, plus a small, approximately spherical, electron-like pocket at the
zone-centre $\Gamma$ point; (d)~a 
``monster''~\cite{Shoenberg}
spanning most of the Brillouin zone; and (e)~a second, electron
approximate
superellipsoid centred on the zone-corner R-point.
As is conventional, the terms ``electron'' and
``hole'' are used for Fermi-surface
sections for which the effective masses are
respectively positive and negative~\cite{Shoenberg}. }
\label{FS}
\end{figure*}

For fixed magnetic field and temperature, the Lifshitz-Kosevich formula~\cite{Shoenberg}
predicts that the amplitude $A$ of a series of de Haas-van Alphen oscillations
should depend on harmonic index $p$ as follows
\begin{equation}
A\propto R_{\rm D}\cos\left(\frac{p\pi g m^*}{2}\right) = {\rm e}^{-\frac{\pi p}{\omega_{\rm c} \tau}}\cos\left(\frac{p\pi g m^*}{2}\right),
\label{dongle}
\end{equation}
where $R_{\rm D}$ is often known as the {\it Dingle factor}.
Here, $\omega_{\rm c} = \frac{eB}{m^*}$ is the cyclotron frequency, $\tau$ is the scattering time,
$B$ is the magnetic flux density~\cite{fieldnote}, $g$ is the $g$-factor
and $m^*$ is the quasiparticle effective mass.
In order to estimate $\tau$ from the harmonic series,
we choose a particular series of dHvA oscillations (fundamental frequency $F=3620$~T)
with a logarithmic amplitude that falls off roughly linearly with $p$
[see Figure~\ref{FFT}(c)].
Such a linear relationship implies that $gm^* \approx 2$ (or $2n$, where $n$ is a nonzero integer),
so that the phase of the cosine term in Equation~\ref{dongle} is approximately constant at each
value of $p$.
Using the effective mass $m^* =0.67 m_{\rm e}$ (see following section),
a fit of the Fourier amplitudes to
Equation~\ref{dongle} for harmonics $p=1-6$ [Figure~\ref{FFT}(c)]
yields $\tau = 0.5\pm 0.1$~ps,
comparable to the scattering times observed in {\it e.g.,} high-purity
copper~\cite{Shoenberg}.
Such values are supported by
the large residual resistivity ratio (RRR) values of the AuBe samples, 
discussed in Section~\ref{RRRhere}.
The value $g\approx 3$ implied by $gm^*\approx 2$ is not unusual
for metals with moderate spin-orbit interactions and electron-electron
interactions~\cite{Shoenberg}. 

The lower-frequency dHvA
oscillations are revealed more clearly by Fourier transforms with a
lower-field window [$4-9$~T; see Figure~\ref{FFT}(d)]~\cite{JdTquery}.
Here, several
frequencies in the range $100-200$~T are observed, with a dominant
peak around 170~T, representing an orbit area that is about 0.90\% of the
square cross-sectional area of the Brillouin zone. 
There are two further peaks between 300 and 350 T,
the higher of which is almost certainly a second harmonic of the 170~T
frequency.

Figure~\ref{FS} shows the calculated bandstructure and Fermi-surface
sections for AuBe inside the simple-cubic Brillouin zone.  The 
bandstructure shown in Figure~\ref{FS}a is similar to that published 
previously~\cite{EteriPirate} and has many features in common with
the bandstructure of other $B20$ materials~\cite{bradlyn,tang,chang,mngecoge}. 
This includes what has previously been described as a fourfold-degenerate 
chiral fermion $\sim 0.4$~eV below the Fermi energy 
at the 
$\Gamma$-point and a chiral double sixfold-degenerate spin-1 Weyl node 
nearly 2~eV below the Fermi level at the R-point of the Brillouin 
zone\cite{tang,chang}. 

The Fermi
surface comprises hole ellipsoids centred on the zone-edge M-points
[Figure~\ref{FS}(b)], two electron approximate superellipsoids centred on the
zone-corner R-points [Figure~\ref{FS}(c, e)], a small, approximately
spherical electron pocket at the zone-centre $\Gamma$ point
[Figure~\ref{FS}(c)] and what old-school fermiologists would call a
{\it monster}~\cite{Shoenberg} of holes [Figure~\ref{FS}(d)].  

For a
polycrystalline sample, the dHvA signal will be
dominated by Fermi-surface pockets that possess either several
extremal orbits that are identical or similar in cross-sectional area,
or a region in which the extremal orbit area varies slowly with
magnetic field orientation~\cite{Shoenberg}.  The predicted Fermi
surface contains several candidate pockets that might cause the
observed dHvA oscillations.  For example, the R-point
pockets sport extremal cross-sectional areas encompassing
the spread of frequencies shown in Figure~\ref{FFT}(b). The largest
and smallest calculated extremal orbits about the M-point ellipsoids are
equivalent to frequencies of 491~T and 1160~T, a range that
again encompasses observed
peaks in the Fourier transforms [Figure\ref{FFT}(e)].  The small $\Gamma$-point pocket is
predicted to have cross-sections corresponding to $F=230-271$~T,
somewhat larger than the pocket suggested by the dominant 
low-frequency oscillation with $F=170$~T.
We will return to a more detailed summary of these attributions in
the discussion of Figure~\ref{MvsF}.

\subsection{Quasiparticle effective masses}
In allocating the various series of dHvA
oscillations to the predicted Fermi-surface sections,
it is useful to examine the
relationship of their effective masses to their frequencies.  The
effective mass $m^*$ given by a dHvA experiment is
defined by~\cite{Shoenberg,Encyc}
\begin{equation}
m^* = \frac{\hbar^2}{2 \pi} \frac{\partial S}{\partial E}
\end{equation}
where $S$ is the $k$-space cross-sectional area of the extremal orbit
in the plane perpendicular to the magnetic field and $E$ is the
quasiparticle energy.  Through the Onsager relationship 
$F =\frac{\hbar}{2 \pi e}S$, the dHvA frequency $F$ is
directly proportional to $S$~[\onlinecite{Shoenberg}].  Therefore, if several
series of dHvA oscillations, corresponding to different
orientations of the magnetic field with respect to the crystal axes,
are derived from one particular band, and this band has a dispersion
relationship $E({\bf k})$ close to the Fermi energy $E_{\rm F}$, then a plot of
$m^*$ versus $F$ will lie on a curve that is characteristic of 
$E({\bf k})$.  As a simple example, 
suppose that the dispersion relationship close to $E_{\rm F}$
is described by $E\propto k^r$,
where $r$ is a constant, 
then $F \propto S \propto E^{\frac{2}{r}}$
and $m^*\propto E^{\frac{2}{r}-1}$ .
Thus, a plot of $m^*$ versus $F$ will have a form
determined by the dispersion relationship
(see Ref.~\onlinecite{Hartstein} and
references therein). 

\begin{figure}[htbp]
\hspace{10mm}
\centering
\includegraphics [width=0.98\columnwidth, angle =0 ]{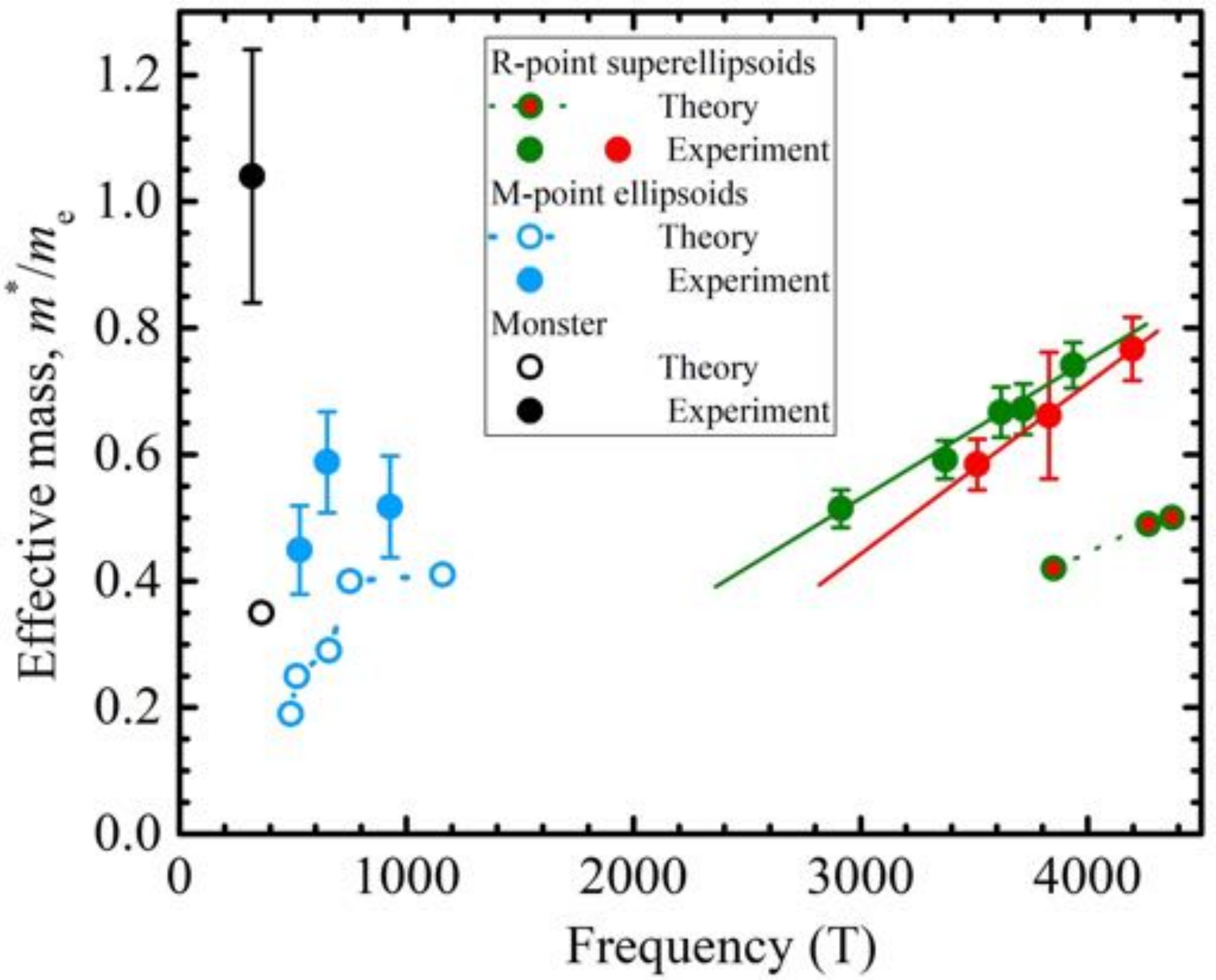}
\caption{Solid points represent effective masses derived using Equation~\ref{LKM} versus
corresponding experimental dHvA frequencies. The error bars are the
uncertainties given by the Simplex fitting routine and the different
coloured points and guide lines are explained in the text.
Masses and frequencies corresponding to a selection of
predicted extremal orbits from the bandstructure
calculations (including spin-orbit interactions) are shown as hollow points.
The hollow light-blue points (calculated for ${\bf H} || [100]$;
dotted line is a guide to the eye)
correspond to the hole
ellipsoids centred on the zone-edge M-points [Figure~\ref{FS}(b)].
The hollow green, red-filled points 
(calculated for ${\bf H} || [1 0 0 ]$)
are from the two electron approximate 
superellipsoids centred on the
zone-corner R-points [Figure~\ref{FS}(c, e)].
The black, hollow point corresponds to a  possible ``monster''
extremal orbit [Figure~\ref{FS}(d)].}
\label{MvsF}
\end{figure}

Figure~\ref{FFT} shows that dHvA effect is observable
over a wide range of temperature $T$, so that the Fourier amplitudes
$A(T, B_{\rm m})$ can be fitted to the temperature-dependent part of
the Lifshitz-Kosevich formula~\cite{Shoenberg}
\begin{equation}
\frac{A(T,B_{\rm m})}{T} \propto \frac{(14.69m^*/B_{\rm m})}{\sinh(14.69m^*T/B_{\rm m})},
\label{LKM}
\end{equation}
where $B_{\rm m}$ is the inverse-field midpoint of the field window used for the Fourier transform:
\begin{equation}
B_{\rm m} = \left[\frac{1}{2} \left(\frac{1}{B_{\rm l}}+\frac{1}{B_{\rm u}}\right)\right]^{-1}.
\label{LKM2}
\end{equation}
Here, $B_{\rm l}$ and $B_{\rm u}$ are, respectively, the lower and
upper field limits of the window.  

Amplitudes from the various series of dHvA oscillations were
fitted numerically to Equation~\ref{LKM} [Figure~\ref{FFT}(f)].  The
$F \approx 170$~T series of dHvA oscillations exhibits a
field-dependent effective mass and will be discussed in more detail
below. The other dHvA frequencies behave in a more
conventional manner, yielding masses that are field-independent within
experimental precision.
 
Experimental effective masses are plotted against their corresponding 
dHvA frequencies in Figure~\ref{MvsF} as solid points. 
The predicted extremal orbits
and the magnitudes of the cyclotron
effective masses from the bandstucture calculations
(including spin-orbit interactions)
are shown as hollow points on the same diagram.
The cluster of frequencies
(solid green and red points in Figure~\ref{MvsF}) spanning the range 
$F=2900- 4200$~T [Figure~\ref{FFT}(a, b)] possess masses from $0.4m_{\rm
e}$ to $0.8 m_{\rm e}$, with $m^*$ approximately proportional to
$F$. It is possible that these masses in fact lie on two separate
curves (indicated by the red and green lines in Figure~\ref{MvsF}).
This suggests that all of these dHvA frequencies are
derived from one or, if one believes the separate green and red
curves, two bands of similar $E$ versus $k$ curvature. 
The theoretical  $m^*, F$ values (green, red-filled, hollow points;  
calculated for ${\bf H} || [1 0 0 ]$)
for the two electron approximate 
superellipsoids centred on the
zone-corner R-points [Figure~\ref{FS}(c, e)]
vary in a similar way 
to the cluster of experimental points, albeit with lower
effective masses.  
These Fermi-surface sections are therefore
very likely to be responsible for the band of dHvA
frequencies from $F= 2900-4200$~T seen in experiments.

The solid light-blue (experimental) points in Figure~\ref{MvsF} group 
around a mass of
$0.5~m_{\rm e}$ and possess frequencies spanning $530-930$~T [Figure~\ref{FFT}(e)].
The predicted $m^*, F$ values (hollow light-blue points; 
calculated for ${\bf H} || [100]$) for extremal orbits of the hole
ellipsoids centred on the zone-edge M-points [Figure~\ref{FS}(b)]
follow a very similar pattern;
it is therefore likely that these Fermi-surface pockets are
responsible for the experimental
dHvA frequencies shown as light-blue points.
As in the case of the R-point Fermi-surface sections
the calculation underestimates the $m^*$ values.

The solid black point in Figure~\ref{MvsF} possesses a low frequency
($F=320$~T) but a substantially higher mass compared to the other oscillation
series observed in the experiments. 
It is possible that this frequency corresponds to one of the
cross-sections of the ``monster'' [Figure~\ref{FS}(d)].
The bandstructure calculations yield several extremal
orbits about the monster, both electron- and hole-like,
with dHvA frequencies $F$
ranging from 181~T to 790~T and effective
mass magnitudes from 0.18 to $0.77m_{\rm e}$.
One of these (black, hollow point) has $F=361$~T, and is a promising 
candidate, possessing an extremal orbit area that varies relatively
slowly with angle.
Just as in the other instances, the effective mass is underestimated
by the calculation, in this case substantially.

\begin{figure}[htbp]
\centering
\includegraphics [width=0.95\columnwidth]{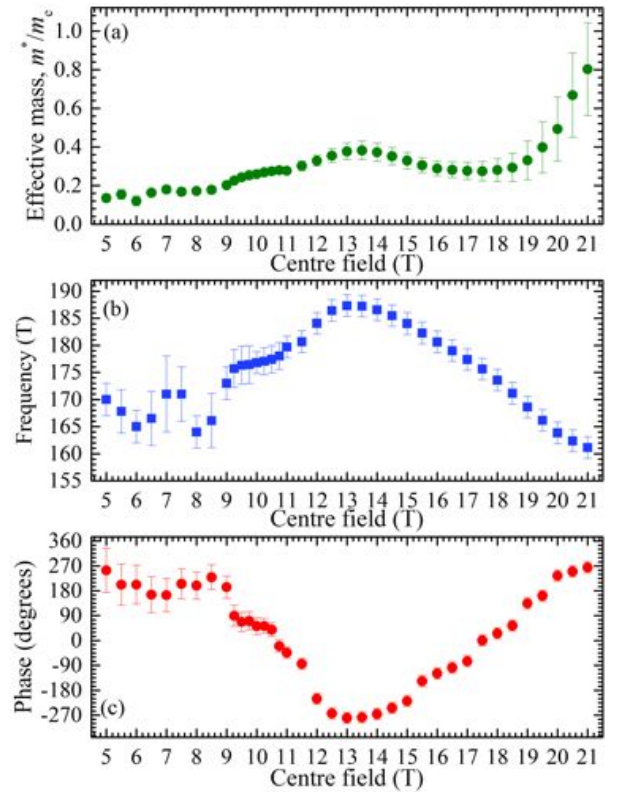}
\caption{Magnetic-field dependence of parameters describing the
$F\approx 170$~T series of dHvA oscillations:
(a)~effective mass $m^*$, (b) frequency $F$ and (c)~phase $\aleph$
(see Equation~\ref{BorisResigns}) of the oscillations.  The field
dependences of these parameters are derived from Fourier analysis of
${\rm d}M/{\rm d} H$ data using inverse field windows of
$\frac{1}{24}$~T$^{-1}$ width symmetrically disposed (in inverse
field) about the centre field ($B_{\rm m})$ values given on the lower
axis.}
\label{EffMass170T}
\end{figure}
The behaviour of the $F \approx 170$~T series of dHvA
oscillations is more unusual.  As noted above, this frequency likely
results from the approximately spherical pocket at the zone-centre
$\Gamma$ point [Figure~\ref{FS}(c)]. Our bandstructure calculations
suggest that this band possesses a linear $E$ versus $k$ (Dirac-like)
dispersion relationship, with the Dirac point~\cite{Wolf} some 400~meV
below the Fermi energy.  The corresponding dHvA
oscillations possess a phase that is difficult to track as a function
of field, plus an effective mass that is field dependent.  In order to
investigate these effects, Fourier transforms of ${\rm d} M/ {\rm d}
H$ data are taken with lower and upper limits given by
\begin{equation}
B_{\rm l} = \left(\frac{1}{B_{\rm m}}+\frac{\Delta}{2}\right)^{-1}
{\rm ~~and ~~}
B_{\rm u} = \left(\frac{1}{B_{\rm m}}-\frac{\Delta}{2}\right)^{-1},
\end{equation}
where the inverse field window $\Delta$ was kept constant at
$1/24$~T$^{-1}$ and the centre field $B_{\rm m}$ was varied in steps
of 0.25~T or 0.5~T.  As well as the amplitude and frequency of each
oscillation series, the Fourier-transform routine produces the phase
$\aleph$ of each oscillation series, defined as
\begin{equation}
\left(\frac{{\rm d} M}{{\rm d}H}\right)_{{\rm osc},F} \propto \cos
\left( 2\pi \frac{F}{B} +\aleph \right),
\label{BorisResigns}
\end{equation}
where the subscript ``osc,~$F$'' indicates the oscillatory component
of the susceptibility associated with the dHvA
oscillations of frequency $F$.  The Fourier amplitudes from different
temperature data can be used to derive the effective mass for each
value of the centre field (see Equation~\ref{LKM}).
To ensure that the mass values were not over-estimates,
only data for temperatures at which the Fourier amplitude was well above the 
noise floor were used in these fits~\cite{mercure}.

To exclude the possibility that the effects under discussion are
due to a superposition of several dHvA frequencies close to 170~T with different masses
and scattering times, several precautions were taken.
(i)~The field window used for the Fourier transforms was wide enough so that any movement, 
growth or change in shape/width of the Fourier peaks due to the emergence of another dHvA series
would be well resolved. Thus, the appearence of a slightly different dHvA frequency as the field or temperature 
changed would be noticeable as an alteration in the width or shape of the peak under study.
Careful observations were made to ensure that this did not occur; {\it e.g.,} as the frequency of the 170~T series 
shifted as the Fourier window moved to higher fields, checks were made to see that there 
was no peak left behind at the original frequency. 
(ii)~The presence of similar frequencies with different masses and phases would tend to 
lead to a shift in frequency and/or phase as the heavier-mass series died 
away with increasing temperature, leaving the lighter-mass oscillations behind.
This was excluded by ensuring that the detected frequency and phase of the 170~T series
remained the same for a particular 
field window as the temperature varied.
(iii)~Finally, high-field (65 T) shots and varying Fourier window widths were used to see whether multiple 
peaks emerged around 170 T; this did not occur, suggesting that the 170 T series is alone.

The results of the above procedure are plotted in Figure~\ref{EffMass170T}.
As can be seen, the effective mass, frequency and phase of the
$F\approx 170$~T dHvA oscillations all vary with
magnetic field, with the mass showing a gradual, but quite spectacular
(factor 5) increase.  
Possible causes for such an effect include a
field-induced change in the energy of the corresponding band relative
to the Fermi energy, a field-induced change in its curvature, 
a field-induced alteration of the many-body effects contributing to
the quasiparticle effective mass, or a combination of all three.
All of these possible effects would 
alter the distribution of quasiparticles between this band and the
reservoir provided by the other Fermi-surface pockets, resulting in
changes to either or {\it both} the dHvA frequency and
the phase of the oscillations~\cite{phasecomment}. 
However, the changes that we measure to the frequency, effective mass, and phase
of the $F\approx 170$~T dHvA oscillations are starkly atypical. 
We are aware of only a small number of materials that exhibit 
a subset of these features, such as a field-induced change (increase or decrease) in 
effective mass in metallic systems associated with Kondo-like phenomena~\cite{HoNJP,cein3sebastian}, 
a class of materials that does not include AuBe. 
However, we point out that the Dirac-like band [Figure~\ref{FS}(a)] that we have
associated with this series 
of dHvA oscillations has been predicted, in this class of 
materials, to undergo a drastic change with 
the application of magnetic field~\cite{tang}. Here, the quasiparticles 
associated with this 
band are expected to split into multiple Weyl fermions due to a breaking 
of time-reversal symmetry.
This unusual modification would have 
a dramatic effect on the dHvA oscillation 
frequency and phase, as well as
the effective mass. Our data suggest that the characteristic field for 
the changes to be experimentally observable is approximately 9~T.
Below this field, the oscillations in ${\rm d}M/{\rm d}H$ have a roughly 
constant phase, similar to that
expected for the Fermi surface of a conventional metal~\cite{Shoenberg}.
Above 9~T, the phase starts to vary, in places exceeding the
extra $\pi$ Berry phase associated with a single species of Weyl fermion~\cite{tang}.
(This shift in phase
of the oscillations with field is the probable reason for an earlier lack of
success in determining an unambiguous Berry phase associated with the
Dirac-like dispersion relationship~\cite{Rebar}.) 

\begin{figure}[htbp]
\vspace{-5mm}
\centering
\includegraphics [width=0.95\columnwidth]{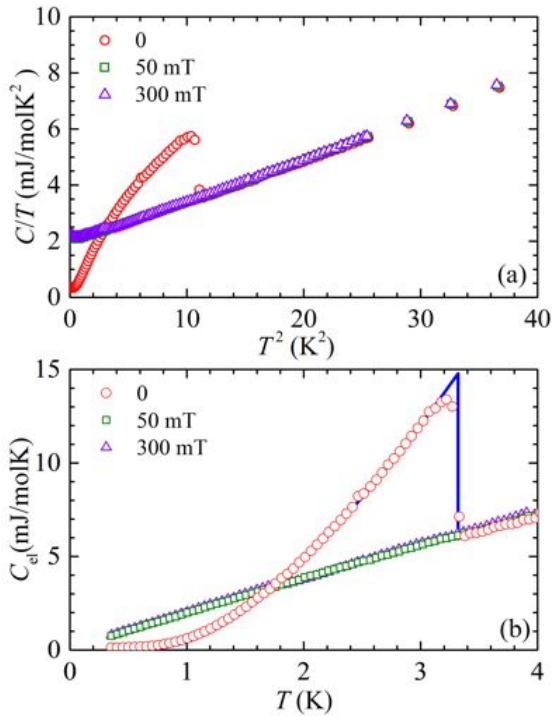}
\vspace{-10mm}
\caption{(a) Heat capacity $C$ of AuBe, divided by temperature $T$,
plotted as a function of $T^{2}$ at fields $\mu_0 H=0$, 50, and
300~mT. The transition to the superconducting state is marked by the
near-vertical jump close to $T^2 = 10$~K$^2$ in the $H=0$ data set.
(b)~The low-temperature electronic component of the heat
capacity versus $T$ at fields of $\mu_0 H=0$, 50, and 300~mT. The
curve is a fit to the standard BCS model for an isotropic,
fully-gapped superconductor in the weak-coupling
limit~\cite{Muhl:numBCS}.}
\label{HeatCap}
\end{figure}

\subsection{Heat capacity}
The low temperature heat capacity $C$ of AuBe is displayed in the form
$C/T$ versus $T^2$ in Figure~\ref{HeatCap}(a).  Here, the
superconducting transition is visible as a jump close to
$T^2=10$~K$^2$ in the $H=0$ data set.  The normal state is restored by
fields of a few tens of mT (see Section~\ref{Sup}), yielding the expected
linear relationship~\cite{AandM}
\begin{equation}
\frac{C}{T} = \gamma +\beta T^2,
\label{MrsMay}
\end{equation}
where $\gamma$ and $\beta$ represent electronic and phonon
contributions respectively.  Straight-line fits of the $H=0$ data set
for temperatures above the superconducting transition and to the
$\mu_0 H=50$ and 300~mT data yield $\gamma = 1.85 \pm 0.06~{\rm
mJ~mol^{-1}K^{-2}}$ and 
$\beta = (1.51 \pm 0.05)\times 10^{-4}~{\rm J~mol^{-1} K^{-4}}$.  
The value of
$\gamma$ is in good agreement with previous
studies~\cite{EteriPirate,Rebar}.
Using our
calculated electronic density of states at the Fermi energy,
$g(E_{\rm F}) \approx 0.59~{\rm states~eV^{-1}(formula~unit)^{-1}}$ 
and the equation~\cite{AandM}
\begin{equation}
\gamma = \frac{\pi^2}{3}k^2_{\rm B} g(E_{\rm F}),
\end{equation}
we obtain a theoretical value 
$\gamma = 1.39~{\rm mJ~mol^{-1}K^{-2}}$,
around 25\% lower than the experimental value.
In a similar way, the bandstructure
calculation underestimates the effective masses
compared to those determined from the temperature dependence of the
dHvA oscillations
[see Figure~\ref{MvsF}].
This suggests that many-body effects
not included in our calculation
contribute to the $g(E_{\rm F})$ and $m^*$
values observed experimentally~\cite{Encyc,AandM}.

The low-temperature phonon contribution $C_{\rm phonon}$ to the heat capacity is
\begin{equation}
C_{\rm phonon}= \frac{12 \pi^4 N k_{\rm B}}{5}\left(
\frac{T}{\theta_{\rm D}}\right)^3 \equiv \beta T^3,
\label{Debye}
\end{equation}
where $\theta_{\rm D}$ is the Debye temperature~\cite{AandM} and $N$
is the number of atoms per formula-mole. For AuBe, $N=2N_{\rm A}$,
where $N_{\rm A}$ is Avogadro's number.  Hence the above value of
$\beta$ can be used to derive $\theta_{\rm D} = 295\pm 3$~K for
AuBe~[\onlinecite{CarelessGermans}]. 

\section{The superconducting state}
\label{Sup}
\subsection{Heat capacity discontinuity}
Figure~\ref{HeatCap}(b) shows the electronic heat capacity $C_{\rm
el}$ obtained by subtracting the phonon contribution,
$\beta$\emph{T}$^{3}$, from $C$, using the value of $\beta$ determined
above.  The superconducting transition in zero applied field is
observed at $T_{\rm c} = 3.3\pm 0.1$~K, with no indication of
superconductivity in fields $\mu_0 H \geq 50$~mT.  The BCS form of the
electronic heat capacity from the work of M\"{u}hlschlegel
\cite{Muhl:numBCS} fitted the data below the superconducting
transition very well.  The ratio of the jump in heat capacity $\Delta
C$ at $T_{\rm c}$ and the normal-state electronic heat capacity
$C_{\rm n}$ above the transition was calculated to be $\Delta C/C_{\rm
n}= 1.48$, a value close to the expected BCS value of 1.43.  Thus, the
heat capacity of AuBe is well described by the standard model for an
isotropic, fully-gapped superconductor in the weak-coupling
limit~\cite{Muhl:numBCS}.
\subsection{Field-temperature phase boundary}
\label{RRRhere}
\begin{figure}[htbp]
\centering
\includegraphics [width=0.8\columnwidth, angle =270 ]{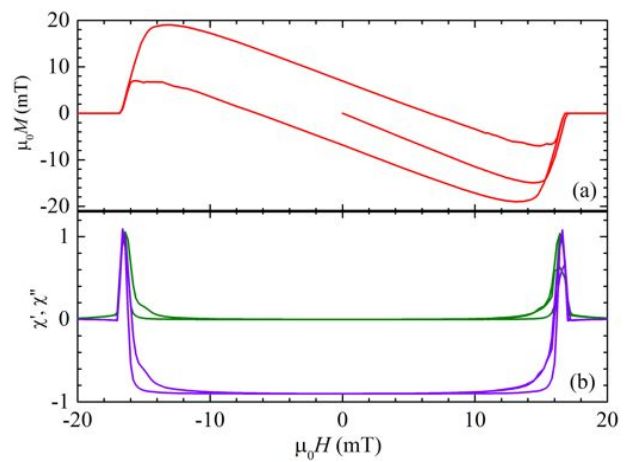}
\vspace{-5mm}
\caption{(a) The dc magnetization $M $(red) versus field $H$ at a
temperature of $T=1.8$~K.  Sharp transitions and a small supercooling,
evident in the hysteresis of the critical field, are apparent
in $M(H)$. (b)~The real component $\chi'$ (purple) and the imaginary
component $\chi''$ (green) of the ac magnetic susceptibility versus
$H$, again at $T=1.8$~K.}
\label{Loopy}
\end{figure}
Figure~\ref{Loopy} shows shows the dc magnetization and ac magnetic susceptibility 
as a function of magnetic field at $T = 1.8$~K,
where the field has been taken through a complete 0 to 40 to $-40$ to 40~mT field cycle. 
The dc magnetization is that of a type I superconductor with
sharp transitions at the critical field $H_{\rm c}$, a small level of
supercooling evident in the hysteresis of the critical
field~\cite{Tinkham}, and a slope in the superconducting state of
$({\rm d}M/{\rm d}H) = -1.1$. This slope is compatible with a
demagnetization factor of $N\approx 0.2$ for the bar-shaped
sample [aspect ratio = length/(square cross sectional 
side)$\approx 2$]~[\onlinecite{Aharo:demag}] and suggests a Meissner effect of -0.90, 
a figure close to the
expectations~\cite{AandM,Tinkham} for a full Meissner effect of
$\chi=-1$. 
Additionally,
measurements on an indium sample of similar dimensions resulted in a
nearly identical Meissner effect, confirming the deduced value of $\chi$. 

The ac magnetic susceptibility in Figure~\ref{Loopy}(b) has been
corrected for demagnetizing effects.  The real component of the ac
susceptibility ($\chi'$) displays a sharp peak at approximately the
midpoint of the superconducting transition.  This peak, known as the
differential paramagnetic effect (DPE) occurs in the intermediate
state ($(1-N)H_{\rm c} \le H \le H_{\rm c}$) and indicates a sudden
expulsion or inclusion of magnetic flux consistent with Type I\cite{leng} or,
rarely, soft Type II superconductivity~\cite{StJames:surfSC,hein}.
Therefore, the superconducting behavior of AuBe near $T_{\rm c}$
indicates a typical Type I response to applied
field~\cite{StJames:surfSC,Tinkham,hein,zhao}.

\begin{figure}[htbp]
\centering
\includegraphics [width=0.85\columnwidth, angle = 0 ]{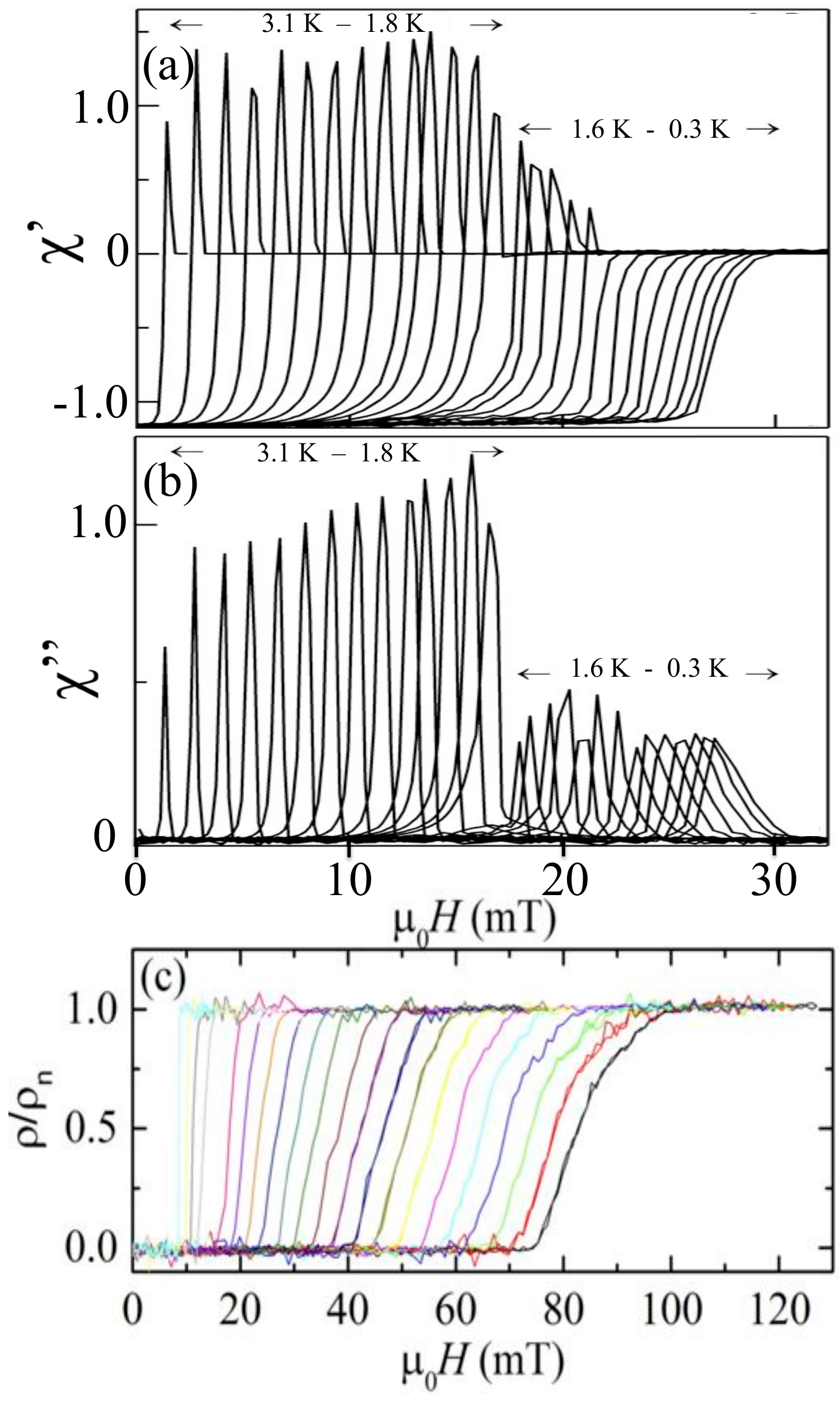}
\vspace{0mm}
\caption{Real component, $\chi'$~(a), and imaginary component,
$\chi''$~(b) of the magnetic susceptibility of a polycrystalline bar
of AuBe plotted against $H$ for a series of constant temperatures in the range
$3.1~{\rm K}\geq T\geq 0.3~{\rm K}$.  
Field sweeps in the temperature range $0.3-1.6$~K were done
in the $^3$He cryostat, while those in the range $1.8-3.1$~K
employed the MPMS.  The jump in peak magnitudes between 1.6~K and
1.8~K is caused by a difference in drive amplitude. (c)~Isothermal
resistivity data, shown as resistivity, $\rho$, divided by the
low-temperature normal-state resistivity, $\rho_{\rm n}$, versus
magnetic~field~$H$. The data were recorded at a series of
approximately equally spaced temperatures covering the range
$3.0~{\rm K} \geq T \geq 0.3$~K.}
\label{Hc2MandR}
\end{figure}

Figure~\ref{Hc2MandR}(a,b) displays the temperature dependence of the
ac magnetic susceptibility down to 0.3~K.  An interesting feature of
these data is the suppression, and eventual disappearance of the DPE
peak in the real component, $\chi'$, as the sample is cooled below
1.2~K.  This is accompanied by a significant increase in the width of
the superconducting transition (in both $\chi '$ and $\chi ''$) below
1.2~K.  The loss of the DPE peak and widening of the transition to the
normal state suggest a crossover into the Type II superconducting
regime~\cite{Annett}.  Therefore, at low temperatures, AuBe seems to
be a Type II superconductor with a Ginzburg-Landau parameter $\kappa
=\lambda/\xi$, where $\lambda$ is the London penetration depth and
$\xi$ is the coherence length, not much larger than $1/\sqrt{2}$, so
that warming above 1.2~K leads to a transition to Type I
behavior~\cite{Tinkham,Annett}.  Similar behavior in disordered
elemental superconductors was investigated thoroughly decades ago and
was termed {\it Type 1.5} or {\it Type II/1 
superconductivity}~\cite{Tinkham,Krag:crossSC,Auer:crossSC} ($\kappa\approx
1/\sqrt{2}$), though it should be noted that the former term has more
recently taken on a new meaning~\cite{egor}. Recently, a 
similar suppression of the DPE in PdTe$_2$ with cooling below 
1.5~K was attributed to screening via a superconducting surface layer~\cite{leng}.
We cannot rule out such a mechanism for AuBe.

The resistivity of AuBe was measured from 0.3~K to room temperature
with a focus on the transition between the normal and superconducting
states.  Although our samples are polycrystalline, having been
synthesized via arc melting, and despite the presence of a small
density of other phases we find resistivities, $\rho$, as low as
$0.2~\mu\Omega\cdot$cm at 4~K and a residual resistivity ratio, RRR =
$\rho {\rm (300~K)}/\rho {\rm (4~K)} = 80$, commensurate
with the large-amplitude dHvA oscillations
discussed in earlier sections.  Figure~\ref{Hc2MandR}(c)
displays the resistivity, $\rho$, divided by the low-temperature
normal-state resistivity, $\rho_{\rm n}$, as a function of applied
field for $T<T_{\rm c}$.  While superconducting transitions are
consistent with critical values and widths determined from the
magnetic characterization for $T > 2.4$~K [compare
Figure~\ref{Hc2MandR}(a,b)], the critical fields for $T < 2.4$~K are
significantly higher than those deduced from $M(H)$, or expected for a
surface state from Ginzburg-Landau theory~\cite{Tinkham,Annett} ($H_{\rm c3} \approx
1.7H_{\rm c2}$).  In agreement with the
broadening of the transition widths found in $\chi(H)$ below 1.2~K in
Figure~\ref{Hc2MandR}(a,b), the resistivity transitions broaden in a
similar fashion at the proposed crossover from Type I to Type II
superconductivity.

To further elucidate the nature of the enhanced critical field in
$\rho$ and compare the behavior of AuBe at fields above $H_{c2}$ to
that of a more standard Type II superconductor hosting a
superconducting surface sheath, a Cr film of thickness between 5 and
10~nm was deposited on the surface of two AuBe samples.  In this way,
a pair-breaking magnetic material~\cite{Annett} has been introduced on all surfaces
of the samples.  Both trials saw no reduction in the critical fields
as determined from $\rho(H)$ measurements.  Thus, resistivity
measurement and subsequent Cr depositions revealed enhanced field
critical values at low temperatures that result from either a surface
state that is insensitive to magnetic scattering or, perhaps, a 
filamentary superconductivity in the bulk of AuBe.

\begin{figure}[htbp]
\centering
\includegraphics[width=0.95\columnwidth]{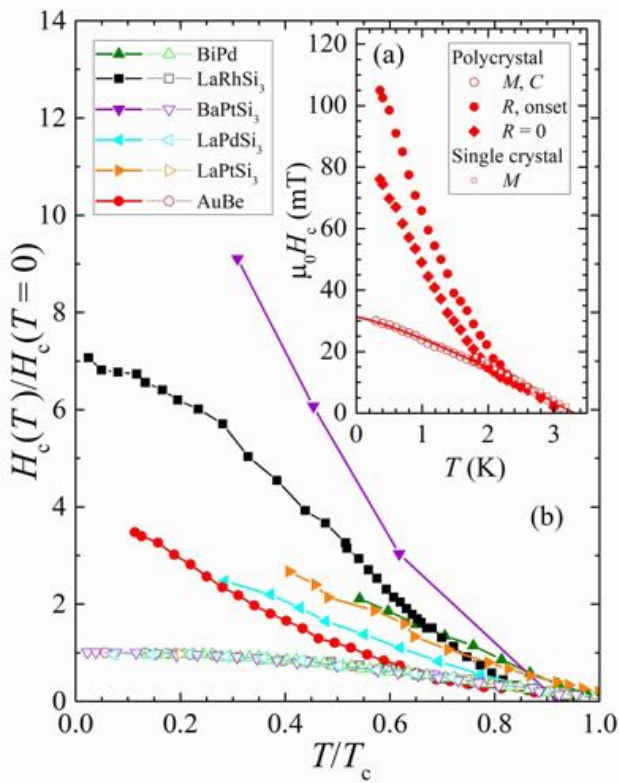}
\caption{(a)~[inset]~Critical fields versus temperature for AuBe. The
solid symbols are critical fields measured using resistivity
[circles = onset of resistive transition, diamonds = highest field
at which zero resistance occurs], 
whilst the hollow symbols represent critical fields derived from
thermodynamic probes (magnetization, susceptibility and heat
capacity). Data from the magnetization of a small single crystal 
are shown as hollow squares. (b)~Analogous phase diagram for a
representative collection of NCSs, presented in reduced
units $H_{\rm c}/H_{\rm c}(T=0)$, $T/T_{\rm c}$.  Here, $H_{\rm
c}(T=0)$ is derived from standard BCS fits to critical fields derived
from thermodynamic probes.  As in (a), open symbols represent
thermodynamic critical fields and filled points are resistive critical
fields.  Data for the various NCSs are from the following sources:
AuBe (this work), BiPd~[\onlinecite{khanthesis}], 
LaRhSi$_{3}$~[\onlinecite{Kimura:LaRhSi}], 
BaPtSi$_{3}$~[\onlinecite{Bauer:BaPtSi}], 
LaPdSi$_{3}$~[\onlinecite{Smid:LaPdSi}], 
LaPtSi$_{3}$~[\onlinecite{Smid:LaPdSi}].}
\label{phaDIA}
\end{figure}

Figure~\ref{phaDIA}(a) displays the superconducting phase diagram for
the critical temperatures $T_{\rm c}$ and critical fields $H_{\rm c}$
determined from what might be termed bulk thermodynamic probes (heat
capacity, magnetization, susceptibility), and resistivity
measurements.  The critical fields and temperatures are defined by the
discernible onset of superconductivity in the particular measurement
technique. (We also include critical fields determined from measurements 
of the magnetization of a small single crystal for comparison to the polycrystalline results. 
Apart from a very slight increase in $T_{\rm c}$,
these data reproduce the data determined from the polycrystalline samples in the 
temperature range explored ($T>1.7$~K).)  
The main result is abundantly clear; the critical field
determined from $\rho(H,T)$, $H_{\rm c \rho}$, diverges from that
determined from $M(H)$ and $\chi'$ at approximately 2.4~K, rising
almost linearly to an extrapolated $T=0$ intercept of approximately
130~mT.  The magnetization critical fields follow a less unusual
variation, yielding
$T_{\rm c}(H=0) = 3.25$~K and $H_{\rm c}(T=0) = 31\pm 1$~mT.
Superconducting fluctuations are not the cause for this large enhancement of 
the resistive critical field, since the critical field determined from the highest 
field displaying zero resistance, shown in Figure~\ref{phaDIA}(a), 
extrapolates to approximately 84~mT at $T=0$. 
This critical field is over 2.6 times larger than $H_{\rm c}(T=0)$.

\section{Discussion}
\label{Disc}

The investigation presented here has permitted a fuller impression 
of the normal and superconducting properties of AuBe, including its electronic structure. 
To place AuBe in context with conventional 
superconductors and other NCSs, we have used these 
experimental data to make estimates
of the most usual parameters used to describe the superconducting state. 
In common with many of the weakly-correlated 
NCSs~\cite{mineev,Agter:NCSbook} and in agreement 
with Reference~[\onlinecite{EteriPirate}],
the fact that the heat capacity of AuBe can be well described by the standard
BCS form for an isotropic superconducting gap with singlet pairing
[Figure~\ref{HeatCap}(b)]  places limits on the
size of the possible triplet contributions to the superconducting
wavefunction. 
Therefore, we use
the BCS expression~\cite{Tinkham}, 
$\Delta = 1.764 k_{\rm B} T_{\rm c}(H=0)$, to derive
$\Delta = 0.49$~meV.  
Furthermore, using the measured dHvA frequencies
({\it i.e.,} Fermi-surface cross-sectional areas) and
effective masses, an average Fermi velocity 
$v_{\rm F} \approx 3.8\times 10^5~{\rm ms}^{-1}$ can be estimated.
Inserting this into the BCS expression~\cite{Tinkham, Annett},
$\xi = \frac{\hbar v_{\rm F}}{\pi \Delta}$,
a coherence length of $\xi \approx 160$~nm results.
Similarly, the calculated bandstructure yields
an averaged effective mass $m^*$ and effective quasiparticle density $n$ that
can be used to estimate the penetration
depth~\cite{Tinkham,Annett},
$\lambda = \sqrt{\frac{m^*}{\mu_0 n q^2}}\approx 120$~nm, where $q$ is 
the electronic charge.
Therefore the Ginzburg-Landau parameter $\kappa
=\lambda/\xi \approx 0.75$, rather consistent with the expectations for
Type II/1 superconductivity~\cite{Tinkham,Krag:crossSC,Auer:crossSC} ($\kappa\approx
1/\sqrt{2}\approx 0.71$) discussed in Section~\ref{Sup}.
A second value of $\kappa$ can be derived from a fit of the BCS parabolic form
of the critical field to the Type II region of the phase diagram~\cite{Tinkham,Annett}.
This gives $\mu_0 H_{\rm c2} = 31$~mT, yielding 
$\kappa = H_{\rm c2}/(\sqrt{2} H_{\rm c})\approx 0.87$,
reasonably close to the value of 0.75 derived from the
estimated $\xi$ and $\lambda$, giving us confidence in our estimates.

The cross-over from Type II (low $T$) to Type
I (higher $T$) superconductivity in AuBe
is similar to behaviour that was investigated decades ago in elemental
superconductors that were intentionally 
disordered~\cite{Krag:crossSC,Auer:crossSC}. 
Though Type I superconductivity is
usually associated with pure elements, and Type II with compounds,
a trend is apparent in that several NCSs display Type I
behaviour~\cite{Smid:LaPdSi,Kimura:2LaRhSi,Anand:LaIrSi,Wakui:RhGa}. 
Since carrier scattering limits the superconducting coherence length, Type I
behavior in AuBe is likely connected to the long mean free path of
charge carriers evident in the very small resistivity found at low
temperature, the observation of dHvA oscillations at 
fields as small as 2~T, and the long scattering time derived from the 
sequence of harmonics in the dHvA oscillations. 
The picture of AuBe that emerges is that of a low-scattering-rate 
NCS that resides near the border between type I and 
II superconductivity. 

Perhaps the most interesting aspect of the superconductivity of AuBe is 
the
enhancement of $H_{\rm c \rho}$ beyond the critical field $H_{\rm c3}$
that would be expected for a superconducting surface
state~\cite{Tinkham}.  This enhancement is unlikely to be due to
defects or to impurity phases residing at surfaces or interfaces of
our samples.  This follows from several important observations
including the correspondence of the critical fields presented in 
Figure~\ref{phaDIA} for $T > 2.4$~K where Type I behavior is apparent. A
superconducting surface state is not expected in materials well into
the Type I regime~\cite{Tinkham} ($\kappa<1/\sqrt{2}$) so that
measurements of the critical fields via magnetic and charge transport
methods coincide.  However, in a superconducting sample with a Type
II-to-Type I crossover, a superconducting surface state would persist
above $H_{\rm c}$ at temperatures somewhat higher than the crossover at
1.2~K, just as we observe for $H_{\rm c \rho}$.  That is, the enhancement of
$H_{\rm c\rho}$ in AuBe is associated with the appearance of Type II
behavior with cooling below $T_{\rm c}$ in a manner similar to that
found for standard superconducting surface states in elemental
superconductors with $\kappa\approx1/\sqrt{2}$.  An enhanced
$H_{\rm c\rho}$ associated with an impurity phase would show no such
connection to the cross-over in character of the bulk superconducting
phase.  In addition, the very small low temperature $\rho$ and
relatively large RRR are not consistent with strong scattering at
interfaces between crystallites such as would likely occur if an
impurity phase was systematically associated with these
interfaces. Thus, we are left with the conclusion that the enhancement
of $H_{\rm c\rho}$ above $H_{c2}$ is likely to be an intrinsic effect
associated with either the surface of our samples, interfaces between
crystallites, or twin boundaries between crystallites of opposite
structural chirality.

This conclusion is reinforced by the reports of an enhanced
$H_{\rm c\rho}$ in several other NCSs including
BiPd~[\onlinecite{khanthesis}], LaRhSi$_{3}$~[\onlinecite{Kimura:LaRhSi,Kimura:2LaRhSi}], 
BaPtSi$_{3}$~[\onlinecite{Bauer:BaPtSi}],
LaPdSi$_{3}$~[\onlinecite{Smid:LaPdSi}], and LaPtSi$_{3}$~[\onlinecite{Smid:LaPdSi}]
making the observation of this effect a trend in such materials. We
have reproduced data from these five NCSs in
Figure~\ref{phaDIA} for comparison to AuBe. The
similarity between LaRhSi$_{3}$ and AuBe is obvious as both of these
materials undergo a transition from Type II-to-Type I behavior at
roughly $0.5\cdot T_{\rm c}$, and both have a region near $T_{\rm c}$ where
$H_{\rm c}$ and $H_{\rm c\rho}$ coincide~\cite{Kimura:LaRhSi}. In contrast BiPd is 
more robustly
type II so that the divergence of $H_{c}$ and $H_{c\rho}$ occurs much
closer to $T_{c}$\cite{khanthesis}. The similarity of the enhanced 
critical fields as
measured in the resistivity in several NCSs
makes problematic sample quality issues less
likely as a cause. We point out that it was common to incorrectly dismiss the
persistence of superconductivity at high field as due to inhomogeneous
samples prior to the discovery of surface superconductivity at fields
as large as $1.695\cdot H_{\rm c2}$ by Saint-James and de 
Gennes~\cite{StJames:surfSC}. However, the cause of the enhanced $H_{\rm c\rho}$
in these NCSs is not clear at this time.  In fact, it
is expected that even in unconventional cases where there is a
spatially-modulated order parameter, the surface superconducting state
will have the same critical field as predicted in Ref.~\onlinecite{StJames:surfSC} 
to first order in $(T_{\rm c} - T)/T_{\rm c}$~[\onlinecite{Agter:NCSbook}]. 
This does leave open the possibility
that higher order terms are responsible for an enhancement of $H_{\rm c3}$
far below $T_{\rm c}$.  Recently, Aoyama {\it et al.} have suggested that
magnetoelectric effects produce an effective magnetic field at twin
boundaries of NCSs which can either enhance or reduce
the local critical fields~\cite{Aoyama:SCtwin}. Since both the
reduction and enhancement are expected at different symmetry grain
boundaries within the same samples, resistivity measurements which are
sensitive to filamentary superconducting pathways would display an
enhanced critical field in this scenario. However, the small size of
the enhancement predicted by these authors is not consistent with the
data displayed in Figure~\ref{phaDIA}.

Because theoretical expectations are not consistent with our data, we
are left to speculate about the reason for an enhanced critical field
in NCSs as measured by the charge carrier transport far
beyond what is expected for a simple surface state in a Type II
superconductor. Of great recent popularity is the idea that
topologically protected superconducting surface states are thought to
form on NCS surfaces or interfaces~\cite{Schnyd:topSurfSC}. 
These interesting states are produced because
of symmetry differences at the sample boundaries. Investigations
searching for topological surface states in several NCSs 
including BiPd were not successful. In BiPd no such states were 
identified 
despite the existence of Dirac-like feature in the surface electronic
structure 0.7~eV below the Fermi energy~\cite{Neu:BiPd,Thiru:BiPd}. The 
enhanced $H_{\rm c\rho}$ in AuBe, as well
as other NCSs displayed in Figure~\ref{phaDIA}, is immune to the
deposition of a magnetic film over its surface, which may be signaling
that novel topologically protected states may be present either
at surfaces or at twin boundaries of opposing chirality. 

Furthermore,
we have shown that AuBe displays a nonstandard field
dependence of the phase of dHvA
oscillations associated with a band thought to 
host unconventional chiral fermions. This result demonstrates the power of dHvA 
techniques to establish the properties of a single
band despite the presence of other electronic bands with a larger density 
of states, even in polycrystalline samples. The existence of unusual bulk
electronic bands suggests the intriguing possibility of equally
unusual, and possibly topological~\cite{latenat}, superconducting surface states with 
enhanced critical fields along with the expectation of Majorana modes. Although we have no
direct evidence for these states, nor a demonstrable connection to the novel
band evident in the electronic structure calculations and the dHvA oscillations, we point out that weakly-correlated
NCSs that have high conductivity such that they display
Type I or 1.5 behavior may be fruitful places to search for these
effects.

\section{Acknowledgments}
JFD acknowledges support through the National Science Foundation (NSF)
grant DMR1206763 for the magnetization and charge transport
measurements as well as polycrystalline synthesis. PWA acknowledges
support through the Department of Energy through DE-FG02-07ER46420 for
specific heat measurements, DPY acknowledges the NSF for support
through DMR1306392 for single crystal synthesis, and JYC the NSF
through DMR1700030 for structural determination through powder X-ray
diffraction. A portion of this work was performed at the National High
Magnetic Field Laboratory, which is supported by the NSF Cooperative
Agreement No. DMR-1157490 and the state of Florida. 
JS thanks the DoE BES FWP {\it Science in 100~T} for supporting the
development of some of the techniques used in this work. We acknowledge
helpful conversations with I. Vekhter and D. Sheehy.



\begin{thebibliography}{99}
\bibitem{Bauer:CePtSi}
E. Bauer, G. Hilscher, H. Michor, C. Paul, E. Scheidt,
A. Gribanov, Y. Seropegin, H. N\''{o}el, M. Sigrist, and P. Rogl, 
Phy. Rev. Lett. {\bf 92}, 027003 (2004).
\bibitem{Yuan:LiPdB}
H. Yuan, D. Agterberg, N. Hayashi, P. Badica, 
D. Vandervelde,
K. Togano, M. Sigrist, and M. Salamon, Phys.
Rev. Lett. {\bf 97}, 017006 (2006).
\bibitem{Schnyd:topSC}
A. P. Schnyder, S. Ryu, A. Furusaki, and A. W. Ludwig,
Phys. Rev. B {\bf 78}, 195125 (2008).
\bibitem{Kaur:helicalSC}
R. Kaur, D. Agterberg, and M. Sigrist, Phys. Rev.
Lett. {\bf 94}, 137002 (2005).
\bibitem{latenat}
Guoqing Chang, Benjamin J. Wieder, Frank Schindler, 
Daniel S. Sanchez, Ilya Belopolski, Shin-Ming Huang, 
Bahadur Singh, Di Wu, Tay-Rong Chang, 
Titus Neupert, Su-Yang Xu, Hsin Lin and M. Zahid Hasan, 
Nature Materials {\bf 17} 978  (2018).
\bibitem{Sun:BiPd}
Z. Sun, M. Enayat, A. Maldonado, C. Lithgow, E. Yelland,
D. C. Peets, A. Yaresko, A. P. Schnyder, and P. Wahl,
Nature Comm. {\bf 6}, 6633 (2015).
\bibitem{khan19}
Mojammel A. Khan, D. E. Graf, I. Vekhter, D. A. Browne, J. F. DiTusa, 
W. Adam Phelan, and D. P. Young,
Phys. Rev. B {\bf 99}, 020507(R) (2019)
\bibitem{Karki:MoAlC}
A. Karki, Y. Xiong, I. Vekhter, D. Browne, P. Adams,
D. Young, K. Thomas, J. Y. Chan, H. Kim, and R. Prozorov,
Phys. Rev. B {\bf 82}, 064512 (2010).
\bibitem{Tsvya:RhGe}
A. Tsvyashchenko, V. Sidorov, A. Petrova, L. Fomicheva,
I. Zibrov, and V. Dmitrienko, J. Alloys and Compounds
{\bf 686}, 431 (2016).
\bibitem{Muhl}
S. M\"{u}hlbauer, B. Binz, F. Jonietz, C. Pfleiderer, 
A. Rosch, A. Neubauer, R. Georgii and P. Boeni,
Science {\bf 323}, 915 (2009).
\bibitem{Matt:AuBe}
B. Matthias, J. Phys. Chem.
Solids {\bf 10}, 342 (1959).
\bibitem{EteriPirate}
A. Amon, E. Svanidze, R. Cardoso-Gil, 
M. N. Wilson, H. Rosner, M. Bobnar, 
W. Schnelle, J. W. Lynn, R. Gumeniuk, 
C. Hennig, G. M. Luke, H. Borrmann, 
A. Leithe-Jasper, and Yu. Grin
Phys. Rev. B {\bf 97}, 
014501 (2018).
\bibitem{Rebar}
Drew Rebar, {\it Exploring superconductivity
in chiral structured AuBe}, Ph.D. dissertation,
Louisiana State University, 2015.
\bibitem{bradlyn} 
B. Bradlyn, J. Cano, Z. Wang, M. G. Vergniory, C. 
Felser, R. J. Cava, and A. Bernevig, Science {\bf 353}, aaf5037 (2016).
\bibitem{tang} 
P. Tang, Q. Zhou, and S.-C. Zhang,
Phys. Rev. Lett. {\bf 119}, 206402 (2017).
\bibitem{chang} 
G. Chang, S.-Y. Xu, B. J. Wieder, D. S. Sanchez,
S.-M. Huang, I. Belopolski, T.-R. Chang, S. Zhang, A. Bansil, H. Lin,
and M. Z. Hasan, Phys. Rev. Lett. {\bf 119}, 206401 (2017).
\bibitem{takane} 
D. Takane, Z. Wang, S. Souma, K. Nakayama, T. Nakamura, 
H. Oinuma, Y. Nakata, H. Iwasawa, C. Cacho, T. Kim, K. Horiba, H. 
Kumigashira, T. Takahashi, Y. Ando, and T. Sato, arXiv:1809.01312v2 (2018).
\bibitem{StJames:surfSC}
D. Saint-James and P. de Gennes, Physics Letters {\bf 7}, 306 (1963).
\bibitem{tiny}
Unfortunately, the tiny single crystal was too small
to give a reliable signal in the apparatus used for
the high-field dHvA experiments~\cite{Rebar}.
\bibitem{HoNJP}
Pei-Chun Ho, J. Singleton, M.B. Maple, 
Hisatomo Harima,
P.A. Goddard, Z. Henkie and A. Pietraszko,
New J. of Physics {\bf 9} 269 (2007).
\bibitem{HystLoops}
John Singleton, Jae Wook Kim, Craig V. Topping, Anders Hansen, Eun-Deok Mun, S. Chikara, 
I. Lakis, Saman Ghannadzadeh, Paul Goddard, Xuan Luo, Yoon Seok Oh, 
Sang-Wook Cheong, and Vivien S. Zapf
Phys. Rev. B {\bf 94}, 224408 (2016).
\bibitem{Blaha:Wien2k}
WIEN2k, an Augmented Plane Wave Plus Local Orbitals
Program for Calculating Crystal Properties (2001).
\bibitem{Perd:GenGrad}
J. P. Perdew, K. Burke, and M. Ernzerhof, Phys. Rev. Lett. {\bf 77}, 16533 (1996).
\bibitem{convnote}
This is a standard benchmark used for density functional theory to test convergence. 
It is the cutoff in the plane-wave basis, {\it i.e.}, the muffin tin radius around each atom, $R$, 
multiplied by the plane-wave cut-off value, $K$.
\bibitem{FeSi}
Pierre Villars (Chief Editor), PAULING FILE in: Inorganic Solid Phases, 
SpringerMaterials (online database), Springer, Heidelberg (ed.) SpringerMaterials 
\begin{verbatim}
(https://materials.springer.com/isp
/crystallographic/docs/sd_1100697 sd_1100697)} 
\end{verbatim}
(Springer-Verlag GmbH, Heidelberg, 2016) 
\bibitem{Cull:AuBe}
B. Cullity, Transactions of the American Institute of Mining
and Metallurgical Engineers 171, 396 (1947).
\bibitem{handbook}
Hiroaki Okamoto 
{\it Desk Handbook 
Phase Diagrams for Binary Alloys} 
(Second Edition), ASM International, Materials Park, 2012.
\bibitem{Greenwood} N.N. Greenwood and A. Earnshaw, {\it Chemistry of
the Elements} (Second Edition). Butterworth-Heinemann, London, 1997.
\bibitem{HeatingComment}
Inductive heating is roughly proportional to 
$({\rm d}H/{\rm d} t)^2$~[\onlinecite{HoNJP}];
the fact that data from the up- and downsweeps of the field,
where ${\rm d}H/{\rm d} t$ values are very different
(see Figure~2 of Ref.~\onlinecite{HystLoops}) overlie
suggests that inductive heating is unimportant.
\bibitem{fieldnote}
The Landau quantization that is
the cause of dHvA oscillations is a result of the
magnetic flux density $B$ experienced by the quasiparticles. In a
non-magnetic material such as AuBe, $B = \mu_0 H$, where $H$ is the
applied magnetic field, is a very good approximation~\cite{Shoenberg},
and so we use $B$ and $\mu_0 H$ reasonably interchangeably.
\bibitem{Shoenberg}
D. Shoenberg, {\it Magnetic Oscillations in Metals}, Cambridge University Press, Cambridge 1984.
\bibitem{AuZn}
P. A. Goddard, J. Singleton, R. D. McDonald, N. Harrison, J. C. Lashley,
H. Harima and M.-T. Suzuki,
Phys. Rev. Lett {\bf 94}, 116401 (2005)
\bibitem{JdTquery}
We Fourier transform the quantity
${\rm d}M/{\rm d}H = ({\rm d}M/{\rm d}t)/({\rm d}H/{\rm d}t)$. 
The voltage from the $\dot{B}$ coil $\propto {\rm d}H/{\rm d}t$ 
tends to zero as the upper limit of the field sweep is approached.
Therefore the Fourier windows are truncated 1 or 2~T below
the maximum field.
\bibitem{mngecoge} 
J. F. DiTusa, S. B.  Zhang, K. Yamaura, Y. Xiong, 
J. C. Prestigiacomo, B. W. Fulfer, P. W. Adams, M. I. Brickson, 
D. A. Browne, C. Capan, Z. Fisk, J. Y. Chan, Phys. Rev. B {\bf 90}, 144404 (2014).
\bibitem{Encyc} 
J. Singleton. {\it Cyclotron Resonance}, in
Encyclopedia of Condensed Matter Physics, Vol. 1, Eds F. Bassani,
G.L. Liedl and P. Wieder, Elsevier, Oxford (2005).
\bibitem{Wolf}
E.L. Wolf, {\it A New Paradigm in Condensed Matter and Device Physics},
Oxford University Press, Oxford, 2013.
\bibitem{Hartstein}
M. Hartstein, W. H. Toews, Y.-T. Hsu, B. Zeng, X. Chen, M. Ciomaga Hatnean, 
Q. R. Zhang, S. Nakamura, A. S. Padgett, G. Rodway-Gant, J. Berk, M. K. Kingston, 
G. H. Zhang, M. K. Chan, S. Yamashita, T. Saraivara, Y. Takano, J.-H. Park, L. Balicas, 
N. Harrison, N. Shitsevalova, G. Balakrishnan, G. G. Lonzarich, R. W. Hill, 
M. Sutherland and S. E. Sebastian,
Nature Physics {\bf 14}, 166 (2018).
\bibitem{mercure}
J.-F. Mercure, A. W. Rost, E. C. T. O’Farrell, S. K. Goh, 
R. S. Perry, M. L. Sutherland, S. A. Grigera, R. A. Borzi, 
P. Gegenwart, A. S. Gibbs, and A. P. Mackenzie
Phys. Rev. B {\bf 81}, 235103 (2010).
\bibitem{phasecomment}
For example, if the energy shift were such that $F = F_0 +\alpha B$,
where $\alpha$ is a constant, the 
apparent frequency of the oscillations
would remain constant, but their phase would shift
(see Equation~\ref{BorisResigns})~\cite{Singleton}.
\bibitem{Singleton}
J. Singleton, R J Nicholas, F Nasir and C K Sarkar, 
J. Phys. C: Solid State Phys., {\bf 19} 35 (1986).
\bibitem{cein3sebastian} 
S. E. Sebastian, N. Harrison, C. D. Batista, S. A. 
Trugman, V. Fanelli, M. Jaime, T. P. Murphy, E. C. Palm, H. Harima,
T. Ebihara, PNAS {\bf 106}, 7741 (2009).
\bibitem{AandM}
N.W. Ashcroft and N.D. Mermin, {\it Solid State Physics},
Holt, Reinhart and Winston, New York, 1976.
\bibitem{CarelessGermans}
We note that the authors of Ref.~\onlinecite{EteriPirate} obtain $\theta_{\rm D}=388$~K.
There is an error in their equivalent of our Equation~\ref{Debye}.
However, this does not seem to be enough to account for the difference in values of $\theta_{\rm D}$.
After numerous careful checks with data from two measurements, we believe that our
value of $\theta_{\rm D}$ is correct.
\bibitem{Muhl:numBCS}
B. M\"{u}hlschlegel, Z. Physik {\bf 155}, 313 (1959).
\bibitem{Tinkham} M. Tinkham, {\it Introduction to
 Superconductivity}, Second Edition (Dover Books on Physics), Vol.~1,
 Dover, New York, 2004.
\bibitem{Aharo:demag}
A. Aharoni, Journal of Applied Physics 83, 3432 (1998).
\bibitem{leng} 
H. Leng, C. Paulsen, Y. K. Huang, and A. de Visser, Phys. Rev. B {\bf 96}, 220506(R) (2017).
\bibitem{hein} 
R. A. Hein and R. L. Falge, Phys. Rev. {\bf 123}, 407
(1961).
\bibitem{zhao} 
L. L. Zhao, S. Lausberg, H. Kim, M. A. Tanatar,
M. Brando, R. Prozorov, and E. Morosan, Phys. Rev. B {\bf 85}, 214526
(2012).
\bibitem{Annett}
 James Annett, {\it Superconductivity, Superfluids and
Condensates}, Oxford University Press, Oxford, 2004.
\bibitem{Krag:crossSC}
U. Kr\"{a}geloh, Phys. Lett. A {\bf 28}, 657 (1969).
\bibitem{Auer:crossSC}
J. Auer and H. Ullmaier, Phys. Rev. B {\bf 7}, 136 (1973).
\bibitem{egor}
E Babaev, J Carlstr\"{o}m, M Silaev, JM Speight,
Physica C: Superconductivity and its Applications {\bf 533}, 20-35 (2017);
{\it Superfluid States of Matter}
(First Edition)
Boris V. Svistunov, Egor S. Babaev, Nikolay V. Prokof'ev,
CRC Press, Boca Raton (2015).
\bibitem{mineev} 
V. P. Mineev and K. V. Samokhin {\it Introduction to 
Unconventional Superconductivity} Gordon and Breach Scientific
Publishers, Amsterdam (1999).
\bibitem{Agter:NCSbook} 
See e.g. {\it Non-centrosymmetric
Superconductors, Introduction and Overview}, Bauer, Ernst, Sigrist,
Manfred (Eds.), Springer, Berlin, 2012, and references therein.
\bibitem{Smid:LaPdSi}
M. Smidman, A. Hillier, D. Adroja, M. Lees, V. Anand,
R. Singh, R. Smith, D. Paul, and G. Balakrishnan, Physical
Review B {\bf 89}, 094509 (2014).
\bibitem{Kimura:2LaRhSi}
N. Kimura, N. Kabeya, K. Saitoh, K. Satoh, H. Ogi,
K. Ohsaki, and H. Aoki, Journal of the Physical Society of
Japan {\bf 85}, 024715 (2016).
\bibitem{Anand:LaIrSi}
V. Anand, D. Britz, A. Bhattacharyya, D. Adroja,
A. Hillier, A. Strydom, W. Kockelmann, B. Rainford, and
K. McEwen, Phys. Rev. B {\bf 90}, 014513 (2014).
\bibitem{Wakui:RhGa}
K. Wakui, S. Akutagawa, N. Kase, K. Kawashima, T. Muranaka,
Y. Iwahori, J. Abe, and J. Akimitsu, J.
Phys. Soc. Japan {\bf 78}, 034710 (2009).
\bibitem{Kimura:LaRhSi}
N. Kimura, H. Ogi, K. Satoh, G. Ohsaki, K. Saitoh,
H. Iida, and H. Aoki, JPS Conf. Proc. {\bf 3}, 015011 (2014).
\bibitem{Bauer:BaPtSi}
E. Bauer, R. Khan, H. Michor, E. Royanian, A. Grytsiv,
N. Melnychenko-Koblyuk, P. Rogl, D. Reith, R. Podloucky,
E.-W. Scheidt, et al., Phys. Rev. B {\bf 80}, 064504 (2009).
\bibitem{Aoyama:SCtwin}
K. Aoyama, L. Savary, and M. Sigrist, Phys. Rev. B
89, 174518 (2014).
\bibitem{Schnyd:topSurfSC}
A. P. Schnyder, P. Brydon, and C. Timm, Phys. Rev.
B {\bf 85}, 024522 (2012).
\bibitem{Neu:BiPd}
M. Neupane, N. Alidoust, M. M. Hosen, J.-X. Zhu, K. Dimitri,
S.-Y. Xu, N. Dhakal, R. Sankar, I. Belopolski, D. S.
Sanchez, et al., Nature Comm {\bf 7}, 13315 (2016).
\bibitem{Thiru:BiPd}
S. Thirupathaiah, S. Ghosh, R. Jha, E. Rienks, K. Dolui,
V. R. Kishore, B. B\"{u}chner, T. Das, V. Awana, D. Sarma,
et al., Phys. Rev. Lett. {\bf 117}, 177001 (2016).
\bibitem{khanthesis} M. Khan, PhD thesis, Louisiana State University,
2017.


\end{thebibliography}
\end{document}